 \documentclass[final,5p,times,twocolumn]{elsarticle}

\usepackage{amssymb}
\usepackage{lipsum}
\usepackage{amsmath}
\usepackage{hyperref}
\usepackage{graphicx}
\usepackage{subcaption}
\usepackage{soul,color,xcolor}
\usepackage{needspace}
\usepackage{booktabs} 
\usepackage{array} 
\usepackage{cuted}

\makeatletter

\journal{Physics Letters B}

\begin{document}

\begin{frontmatter}



\title{Thermodynamics and P-v criticality of RN-AdS black hole surrounded by PFDM on the EGUP framework}


\author{Bo-Li Liu\fnref{a}}

\author{Yu Zhang\corref{cor1}\fnref{a}}
\ead{zhangyu\_128@126.com}
\affiliation[a]{Faculty of Science, Kunming University of Science and Technology, Kunming, Yunnan 650500, China}

\author{Qi-Quan Li\fnref{b}}

\affiliation[b]{School of Physics Science and Technology, Xinjiang University, Urumqi 830046, China}

\cortext[cor1]{Corresponding author}

\begin{abstract}
In this paper, we systematically study the thermodynamic properties and P-v criticality of the RN-AdS black hole surrounded by PFDM in the framework of the extended generalized uncertainty principle (EGUP). In the extended phase space, starting from the first law of black hole thermodynamics, the EGUP-corrected Hawking temperature, heat capacity, and entropy are derived. Then, we use graphical methods to analyze the effects of the EGUP parameters on these thermodynamic quantities. Analysis of the heat capacity indicates that the EGUP correction is beneficial for maintaining the thermodynamic stability of black holes. Finally, based on the modified thermodynamic quantities, we obtain the EGUP-corrected black hole equation of state and Gibbs free energy. The G-T diagram with the EGUP correction effect is analyzed, and the results show that when $P<P_c$, a ``swallowtail'' behavior emerges, indicating that the black hole undergoes a first order phase transition. The position of the phase equilibrium point changes slightly as the EGUP parameters increase. For the cases of $P=P_c$ and $P>P_c$, we combine the analysis of the relationship between heat capacity and event horizon radius, finding that the phase transition characteristics depicted in the G-T diagram are similar to those without the correction effect.

\end{abstract}




\end{frontmatter}




\section{Introduction}\label{introduction}

The establishment of black hole thermodynamics began with the pioneering work of Bekenstein and Hawking in the 1970s. Bekenstein proposed that the entropy of a black hole is proportional to the area of its event horizon, while Hawking used quantum field theory in curved spacetime to show that black holes have temperature and radiate particles \cite{Bekenstein:1973ur,Hawking:1975vcx}. Together, they reveal the nature of black holes as thermodynamic systems. Furthermore, Bardeen, Carter and Hawking proposed the four laws of black hole thermodynamics \cite{Bardeen:1973gs}. Among them, the first law of black hole thermodynamics did not include the "pressure-volume" term, which would make the heat capacity of the black hole negative and show thermodynamic instability. 
Subsequent studies have shown that this problem can be solved in AdS space-time by matching the cosmological constant $\Lambda$, interpreting it as the thermodynamic pressure $P = -\frac{\Lambda}{8\pi}$, and introducing the conjugate volume V \cite{Creighton:1995au,Hubeny:2014bla,Padmanabhan:2002sha}. In this way, the expanded first law of thermodynamics can be obtained, which directly provides a theoretical background for our study of the thermodynamic stability and the phase transition of black holes \cite{Teitelboim:1985dp,Brown:1987dd}. Meanwhile, black holes in AdS spacetime have been widely studied because of their rich phase structures \cite{Sadeghi:2020xtc,Wei:2014hba,Singh:2020xju,Hamil:2023zeb,Luciano:2023bai,Guo:2019oad,Zhang:2024jlp,Wang:2024jtp}.

Astronomical observations show that dark matter accounts for approximately 26.8\% of the total mass-energy of the universe \cite{Planck:2015fie}. This invisible matter dominates the large-scale structure of the universe through gravitational effects. Kiselev \cite{Kiselev:2002dx,Kiselev:2003ah} first proposed the energy–momentum tensor for perfect fluid dark matter (PFDM) and gave a static spherically symmetric solution in the dark matter background. The metric function includes the logarithmic term of the non-zero dark matter contribution $\alpha \ln\left( r / r_q \right)$. He also explained the asymptotic behavior of rotation curves in spiral galaxies based on this black hole solution. The PFDM model has been widely discussed in studies of black hole thermodynamics \cite{Liang:2023jrj,Sood:2024rfr,Hamil:2024neq,Singh:2025svv,Li:2024tyk,Abdusattar:2023xxs,Sekhmani:2025kav}. Xu et al. used the specific cases of the black hole solution given in Refs.~\cite{Kiselev:2002dx} and \cite{Li:2012zx} to obtain the RN-AdS black hole solution under PFDM, and studied the thermodynamics and phase transitions of the black hole in the extended phase space defined by the squared charge $Q^2$ and the conjugate quantity $\psi$ \cite{Xu:2016ylr}. In addition to PFDM, other dark matter candidates that have received attention include weakly interacting massive particles (WIMPs), axions, primordial black holes, and neutral particles.

It is worth noting that black hole thermodynamics is usually studied in the framework of semiclassical theory. When the mass of a black hole approaches the Planck scale, traditional semiclassical methods are no longer applicable, and we need to rely on a complete quantum gravity theory. Research in many fields of physics has confirmed the existence of a minimum physical length, such as quantum gravity \cite{Garay:1994en}, non-commutative geometry \cite{Capozziello:1999wx}, and string theory \cite{Konishi:1989wk}. Near the Planck scale, the Heisenberg uncertainty principle (HUP) needs to be modified to the generalized uncertainty principle (GUP), which supports the existence of a minimum observable length \cite{Nozari:2012gd,Maggiore:1993kv,Ali:2009zq,Menculini:2013ida,Majumder:2011hy,Battista:2024gud}. Additionally, loop quantum gravity and black hole physics have also shown the need for corrections to the HUP \cite{Gross:1987kza,Amati:1988tn,Maggiore:1993rv,Bojowald:2011jd,Scardigli:1999jh,Scardigli:2003kr}. Applying GUP to the study of black hole thermodynamics may provide explanations for some issues in classical black hole thermodynamics. For example, a new GUP was proposed to correct the thermodynamics of a Schwarzschild black hole in Ref.~\cite{Banerjee:2010sd}, and it was found that the evaporation of the black hole would stop at the stage of remnant mass, which helped to avoid the singularity problem of black holes. It should be pointed out that since GUP cannot be applied to spaces with large length scales, Park proposed the extended uncertainty principle (EUP) \cite{Park:2007az}. The linear combination of EUP and GUP can lead to the EGUP \cite{Bolen:2004sq}. Tan used EGUP to analyze and calculate the thermodynamic quantities of an RN-AdS black hole surrounded by quintessence matter \cite{Tan:2024sgv}. There is still much research on applying EGUP to black hole thermodynamics \cite{Wu:2022leh,Chaudhary:2024yod,Chen:2022dap,Farahani:2020mfl,Hamil:2023obc,Hassanabadi:2020osz}. In this paper, we will study how EGUP affects the thermodynamic properties and P-v criticality of the RN-AdS black hole surrounded by ideal fluid dark matter—this is an interesting question worth exploring.

To this end, we have organized the structure of this article as follows: in Sec.~\ref{II}, we briefly review the EGUP formalism. In Sec.~\ref{III}, we apply the EGUP to derive and analyze the thermodynamic quantities of the RN-AdS black hole surrounded by PFDM. Then, in Sec.~\ref{IV}, we study the P-v criticality in the EGUP framework. Finally, we conclude the research results in Sec.~\ref{V}.

\section{Extending the generalized uncertainty principle}\label{II}
With the development of quantum theory, the HUP was modified to the GUP near the Planck scale, a modification that introduces a minimum length scale and can be expressed as \cite{Chen:2022dap,Cavaglia:2004jw}
\begin{equation}\label{eq:E21}
\Delta x_{i} \Delta p_{j} \geq \frac{\hbar}{2}\delta _{ij} \left[1+\frac{\beta l_{p}^{2}}{\hbar^{2}}(\Delta p_{j})^{2}\right].
\end{equation}
Here $x_i$ and $p_j$ represent the spatial coordinates and momenta, respectively, while $\beta$ is the dimensionless GUP parameter of order one, and $l_p$ denotes the Planck length. From Eq.  (\ref{eq:E21}), the range of momentum uncertainty can be derived as follows:
\begin{equation}\label{eq:E22}
\frac{\hbar \Delta x_{i}}{\beta l_{p}^{2}}\left[1-\sqrt{1-\frac{\beta l_{p}^{2}}{(\Delta x_{i})^{2}}}\right] \leq \Delta p_{i} \leq \frac{\hbar \Delta x_{i}}{\beta l_{p}^{2}}\left[1+\sqrt{1-\frac{\beta l_{p}^{2}}{(\Delta x_{i})^{2}}}\right].
\end{equation}
From the above formula, we can conclude that the minimum observable length is $\Delta x \geq \sqrt{\beta} l_{p}$.
It is noteworthy that GUP is no longer applicable for spaces with large length scales. The EUP was proposed in Ref. \cite{Park:2007az}, which takes the form:
\begin{equation}\label{eq:E23}
\Delta x_{i} \Delta p_{j} \geq\frac{ \hbar \delta_{i j}}{2} \left[1+\alpha ^{2} \frac{\left(\Delta x_{i}\right)^{2}}{L^{2}}\right],
\end{equation}
where $\alpha$ is the dimensionless EUP parameter of order one and $L$ signifies the large-scale length. It can be readily shown that the minimum momentum uncertainty, as given by Eq.  (\ref{eq:E23}), is $( \Delta p)_{min} =\hbar \alpha /L$.

The linear combination of EUP and GUP can be obtained as EGUP, which is given by
\begin{equation}\label{eq:E24}
\Delta x_{i} \Delta p_{j} \geq \frac{\hbar }{2}\delta _{ij}\left[1+\beta l_{p}^{2} \frac{(\Delta p_{j})^{2}}{\hbar^{2}}+\alpha  \frac{(\Delta x_{i})^{2}}{L^{2}}\right].
\end{equation}
Accordingly, Eq. (\ref{eq:E24}) yields the ranges of uncertainty in momentum and position:
\begin{equation}\label{eq:E25}
\begin{split}
&\frac{\hbar \Delta x_{i}}{\beta l_{p}^{2}}\left[1-\sqrt{1-\beta l_{p}^{2}\left(\frac{1}{(\Delta x_{i})^{2}}+\frac{\alpha}{L^{2}}\right)}\right] \leq \Delta p_{i}\\& 
\leq \frac{\hbar \Delta x_{i}}{\beta l_{p}^{2}}\left[1+\sqrt{1-\beta l_{p}^{2}\left(\frac{1}{(\Delta x_{i})^{2}}+\frac{\alpha}{L^{2}}\right)}\right],
\end{split}
\end{equation}

\begin{equation}\label{eq:E26}
\begin{split}
&\frac{L^{2} \Delta p_{i}}{\hbar \alpha}\left[1-\sqrt{1-\frac{\alpha \hbar^{2}  }{L^{2}}\left(\frac{1}{(\Delta p_{i})^{2}}+\frac{\beta l_{p}^{2}}{\hbar^{2}} \right)}\right] \leq \Delta x_{i}
 \\& \leq \frac{L^{2} \Delta p_{i}}{\hbar \alpha}\left[1+\sqrt{1-\frac{\alpha \hbar^{2}}{L^{2}}\left(\frac{1}{(\Delta p_{i})^{2}}+\frac{\beta l_{p}^{2}}{\hbar^{2}} \right)}\right].
\end{split}
\end{equation}
Based on Eqs. (\ref{eq:E25}) and (\ref{eq:E26}), the following sections employ this formalism to investigate the thermodynamic properties and phase transitions of the RN-AdS black hole surrounded by the PFDM. Throughout this paper, we adopt the natural unit system, in which $c=G=\hbar=1$.

\section{Thermodynamic properties of RN-AdS black hole surrounded by PFDM in the EGUP framework}\label{III}
When the dark matter field is minimally coupled with gravity, the electromagnetic field, and the cosmological constant, it can be described by the following action \cite{Kiselev:2003ah,Li:2012zx,Xu:2016ylr}:
\begin{equation}\label{eq:E31}
S = \int \mathrm{d}^4 x \sqrt{-g} \left( \frac{R}{16\pi} - \frac{\Lambda}{8\pi} + \frac{F_{\mu\nu} F^{\mu\nu}}{4} + \mathcal{L}_{\mathrm{DM}} \right),
\end{equation}
where $\Lambda$ is the cosmological constant, $\mathcal{L}_{\mathrm{DM}}$ signifies the Lagrangian related to the density of ideal fluid dark matter, and $F_{\mu\nu}$ denotes the electromagnetic tensor.

The line element of the RN-AdS black hole surrounded by PFDM is \cite{Xu:2016ylr,Shaymatov:2020wtj}
\begin{equation}\label{eq:E32}
d s^{2}=-f(r) d t^{2}+\frac{1}{f(r)} d r^{2}+r^{2}\left(d \theta^{2}+\sin ^{2} \theta d \phi^{2}\right),
\end{equation}
where 
\begin{equation}\label{eq:E33}
f(r)=1-\frac{2 M}{r}+\frac{Q^{2}}{r^{2}}+\frac{r^{2}}{l^{2}}+\frac{\lambda}{r} \ln_{}{\frac{r}{|\lambda|}}.
\end{equation}

Among them, $Q$ denotes the black hole charge, $\lambda$ is related to the density and pressure of dark matter, and $l$ represents the AdS space-time radius. In this paper, for the PFDM model, we study the case where $\lambda> 0$, which can avoid negative energy density.\\
When $f(r)=0$, the outer horizon radius of the black hole $r_+ $ can be obtained by
\begin{equation}\label{eq:E34}
\left ( 1-\frac{2 M}{r}+\frac{Q^{2}}{r^{2}}+\frac{r^{2}}{l^{2}}+\frac{\lambda}{r} \ln_{}{\frac{r}{|\lambda|}}  \right )\Bigg| _{r=r_+}=0.
\end{equation}
By solving Eq. (\ref{eq:E34}), we can also deduce the black hole mass $M$ as
\begin{equation}\label{eq:E35}
M=\frac{r_{+}}{2}+\frac{Q^{2}}{2 r_{+}}+\frac{r_{+}^{3}}{2l^{2}}+\frac{\lambda}{2} \ln_{}{\frac{r_{+}}{|\lambda|}}.
\end{equation}
From this, the surface gravity $\kappa$ is given by
\begin{equation}\label{eq:E36}
 \kappa=-\lim _{r \rightarrow r_{+}} \frac{\partial_{r} g_{\mathrm{tt}}}{\sqrt{-g_{\mathrm{tt}} g_{\mathrm{rr}}}}=\left.f(r)^{\prime}\right|_{r=r_{+}}=-\frac{Q^{2}}{r_{+}^{3}}+\frac{1}{r_{+}}+\frac{3 r_{+}}{l^{2}}+\frac{\lambda}{r_{+}^{2}}.
\end{equation}

In the extended phase space, where the black hole mass is understood as enthalpy, the first law of thermodynamics for the black hole takes the form:
\begin{equation}\label{eq:E37}
d M=T d S+\Phi d Q+V d P+\Pi d \lambda
\end{equation}
where $\Pi=\frac{\partial M}{\partial \lambda}=\frac{1}{2}\left[\ln \left(\frac{r_{+}}{|\lambda|}\right)-1\right]$ is the conjugate 
quantity and $\Phi=Q/r_+$ represents the electrostatic potential. Based on the above equations, we now investigate the thermodynamic properties of black holes in the EGUP framework.\\
From Eq. (\ref{eq:E37}) and Ref. \cite{Xiang:2009yq}, the Hawking temperature of a black hole is
\begin{equation}\label{eq:E38}
T=\left(\frac{\partial M}{\partial S}\right)_{Q,P,\lambda}=\frac{d A}{d S} \times\left(\frac{\partial M}{\partial A}\right)_{Q,P,\lambda}=\frac{d A}{d S} \times \frac{\kappa}{8 \pi}
\end{equation}
According to Ref. \cite{Xiang:2009yq}, when a particle is captured by a black hole, the information of the particle may be lost. Information theory shows that the minimum unit of information loss is one bit, which corresponds to the minimum change in entropy being $\ln_{}{2}$. At the same time, this process makes the minimum change in the black hole area proportional to the product of position and momentum uncertainties \cite{Chen:2022dap,Liu:2025eok}. Therefore, the entropy of the black hole and the horizon area have the following relationship:   
\begin{equation}\label{eq:E39}
\frac{d A}{d S} \simeq \frac{(\Delta A)_{\min }}{(\Delta S)_{\min }} \simeq \frac{\gamma}{\ln 2} \Delta X \Delta P .
\end{equation}
Next, substituting Eq. (\ref{eq:E39}) into Eq. (\ref{eq:E38}), we can derive the expression for the Hawking temperature, which reads
\begin{equation}\label{eq:E40}
T \simeq \frac{\kappa \gamma}{8 \pi \ln 2} \Delta X \Delta P
\end{equation}
Here, $\gamma$ is a correction factor to ensure that Eq. (\ref{eq:E40}) can reproduce the semiclassical result. Hassanabadi et al. assume that the relationship between particle size and position uncertainty is $\Delta X \simeq 2 r_{+}$  \cite{Hassanabadi:2019eol}. By combining Eqs. (\ref{eq:E25}) and (\ref{eq:E36}), the corrected Hawking temperature can be written as
\begin{equation}\label{eq:E310}
\begin{split}
T&=\frac{\gamma r_+}{2 \pi \beta  l_{p}^{2} \ln_{}{2} }\left(1-\frac{Q^{2}}{r_{+}^{2}}+\frac{3 r_{+}^{2}}{l^{2}}+\frac{\lambda}{r_{+}}\right)
\\&\times\left(1-\sqrt{1-\beta l_{p}^{2}\left(\frac{1}{4 r_{+}^{2}}+\frac{\alpha}{L^{2}}\right)}\right)
\end{split}
\end{equation}
Obviously, when Eq. (\ref{eq:E310}) satisfies $Q \rightarrow 0, \lambda \rightarrow 0, \alpha \rightarrow 0, \beta \rightarrow 0,l \rightarrow \infty$, we have $T=\gamma/\left ( 16 \pi r_{+} \ln_{}{2} \right ) $. To obtain $T=1/\left ( 4 \pi r_{+} \right ) $, we take $\gamma$ to be $4 \ln_{}{2}$. Thus, the EGUP-corrected Hawking temperature of the RN-AdS black hole surrounded by PFDM can be denoted as
\begin{equation}\label{eq:E311}
\begin{split}
T_{EGUP}&=\frac{2 r_{+}}{\pi \beta l_{p}^{2}}\left(1-\frac{Q^{2}}{r_{+}^{2}}+\frac{3 r_{+}^{2}}{l^{2}}+\frac{\lambda}{r_{+}}\right)
\\&\times \left(1-\sqrt{1-\beta l_{p}^{2}\left(\frac{1}{4 r_{+}^{2}}+\frac{\alpha}{L^{2}}\right)}\right).
\end{split}
\end{equation}
When $\alpha = 0$, the GUP-corrected Hawking temperature reads as follows:
\begin{equation}\label{eq:E312}
T_{GUP}=\frac{2 r_{+}}{\pi \beta l_{p}^{2}}\left(1-\frac{Q^{2}}{r_{+}^{2}}+\frac{3 r_{+}^{2}}{l^{2}}+\frac{\lambda}{r_{+}}\right)\left(1-\sqrt{1-\frac{\beta l_{p}^{2}}{4 r_{+}^{2}}}\right).
\end{equation}
Similarly, when $\beta= 0$, the EUP-corrected Hawking temperature is
\begin{equation}\label{eq:E313}
T_{EUP}=\frac{r_{+}}{\pi}\left(1-\frac{Q^{2}}{r_{+}^{2}}+\frac{3 r_{+}^{2}}{l^{2}}+\frac{\lambda}{r_{+}}\right)\left(\frac{\alpha}{L^{2}}+\frac{1}{4 r_{+}^{2}}\right).
\end{equation}
While for $\alpha = 0$ and $\beta= 0$, Eq. (\ref{eq:E311}) reduces to the HUP-corrected Hawking temperature:
\begin{equation}\label{eq:E314}
T_{HUP}=\frac{1}{4 \pi r_{+}}\left(1-\frac{Q^{2}}{r_{+}^{2}}+\frac{\lambda}{r_{+}}+\frac{3 r_{+}^{2}}{l^{2}}\right)
\end{equation}
It should be pointed out that the Hawking temperature of the black hole must be a positive real value, so we need to constrain the horizon radius. From Eq. (\ref{eq:E311}), it can be seen that there are two constraints, one is determined by the properties of the black hole, and the other by the EGUP parameters. Therefore, we have the following constraints on the horizon radius. 
On the one hand, we have
\begin{equation}\label{eq:E315}
1-\frac{Q^{2}}{r_{+}{ }^{2}}+\frac{3 r_{+}{ }^{2}}{l^{2}}+\frac{\lambda}{r_{+}} \geq 0
\end{equation}
By solving Eq. (\ref{eq:E315}), we obtain the following roots:
\begin{equation}\label{eq:E316}
\begin{split}
r_{+1}&=\pm \frac{1}{2} \sqrt{-\frac{4l^{2}}{9}-\frac{\chi^{1 / 3}}{9 \times 2^{1 / 3}}-\frac{\chi_{0}}{9}-\frac{2l^{2} \lambda}{\sqrt{-2l^{2}+\frac{\chi^{1 / 3}}{2^{1 / 3}}+\chi_{0}}}}
\\&+\frac{1}{6} \sqrt{-2l^{2}+\frac{\chi^{1 / 3}}{2^{1 / 3}}+\chi_{0}},
\end{split}
\end{equation}
\begin{equation}\label{eq:E317}
\begin{split}
r_{+2}&=\pm \frac{1}{2} \sqrt{-\frac{4l^{2}}{9}-\frac{\chi^{1 / 3}}{9 \times 2^{1 / 3}}-\frac{\chi_{0}}{9}+\frac{2l^{2} \lambda}{\sqrt{-2l^{2}+\frac{\chi^{1 / 3}}{2^{1 / 3}}+\chi_{0}}}}
\\&-\frac{1}{6} \sqrt{-2l^{2}+\frac{\chi^{1 / 3}}{2^{1 / 3}}+\chi_{0}},
\end{split}
\end{equation}
where
\begin{equation}\label{eq:E318}
\chi_{0}=\frac{2^{1 / 3}\left(l^{4}-36l^{2} Q^{2}\right)}{\chi^{1 / 3}},
\end{equation}
\begin{equation}\label{eq:E319}
\begin{split}
\chi&=2l^{6}+216l^{4} Q^{2}+81l^{4} \lambda^{2}
\\&+\sqrt{-4\left(l^{4}-36l^{2} Q^{2}\right)^{3}+\left(2l^{6}+216l^{4} Q^{2}+81l^{4} \lambda^{2}\right)^{2}}.
\end{split}
\end{equation}
On the other hand, from the square root term of Eq. (\ref{eq:E311}) we can obtain
\begin{equation}\label{eq:E320}
1-\beta l_{p}^{2}\left(\frac{1}{4 r_{+}^{2}}+\frac{\alpha}{L^{2}}\right) \geq 0,
\end{equation}
\begin{equation}\label{eq:E321}
1-\frac{\beta l_{p}^{2}}{4 r_{+}^{2}} \geq 0.
\end{equation}
The premise of this constraint is that $\beta \ne 0$, so neither HUP nor EUP meets this condition. Thus, we derive the second constraint range of the horizon radius,
\begin{equation}\label{eq:E322}
r_{ EGUP }\ge \frac{l_{p} L}{2} \sqrt{\frac{\beta}{L^{2}-\alpha \beta l_{p}^{2}}},
\end{equation}
\begin{equation}\label{eq:E323}
r_{GUP}\ge\frac{l_{p}}{2} \sqrt{\beta}.
\end{equation}
Based on the above analysis, the horizon radius must satisfy both Eqs. (\ref{eq:E315}) and (\ref{eq:E320}). Considering that $P=\frac{3}{8 \pi l^{2}}$, we take $P = 0.037$, $\lambda = 0.1$, and $Q = 0.3$ to calculate the first constraint range: $r_{+}\ge 0.248$. For the second constraint, we consider four cases with EGUP parameter values: $\alpha$ = $\beta$ = 0.01, 0.03, 0.05, and 0.06. The resulting horizon radius ranges are $r_{+}\ge 0.050$, $r_{+}\ge 0.087$, $r_{+}\ge 0.112$, and $r_{+}\ge 0.123$, respectively. Therefore, we take the intersection of the ranges under the two constraints and find that the range of the horizon radius is 
\begin{equation}\label{eq:E324}
r_{+}\ge 0.248.
\end{equation}

\begin{figure}
	\centering 
	\includegraphics[width=0.45\textwidth]{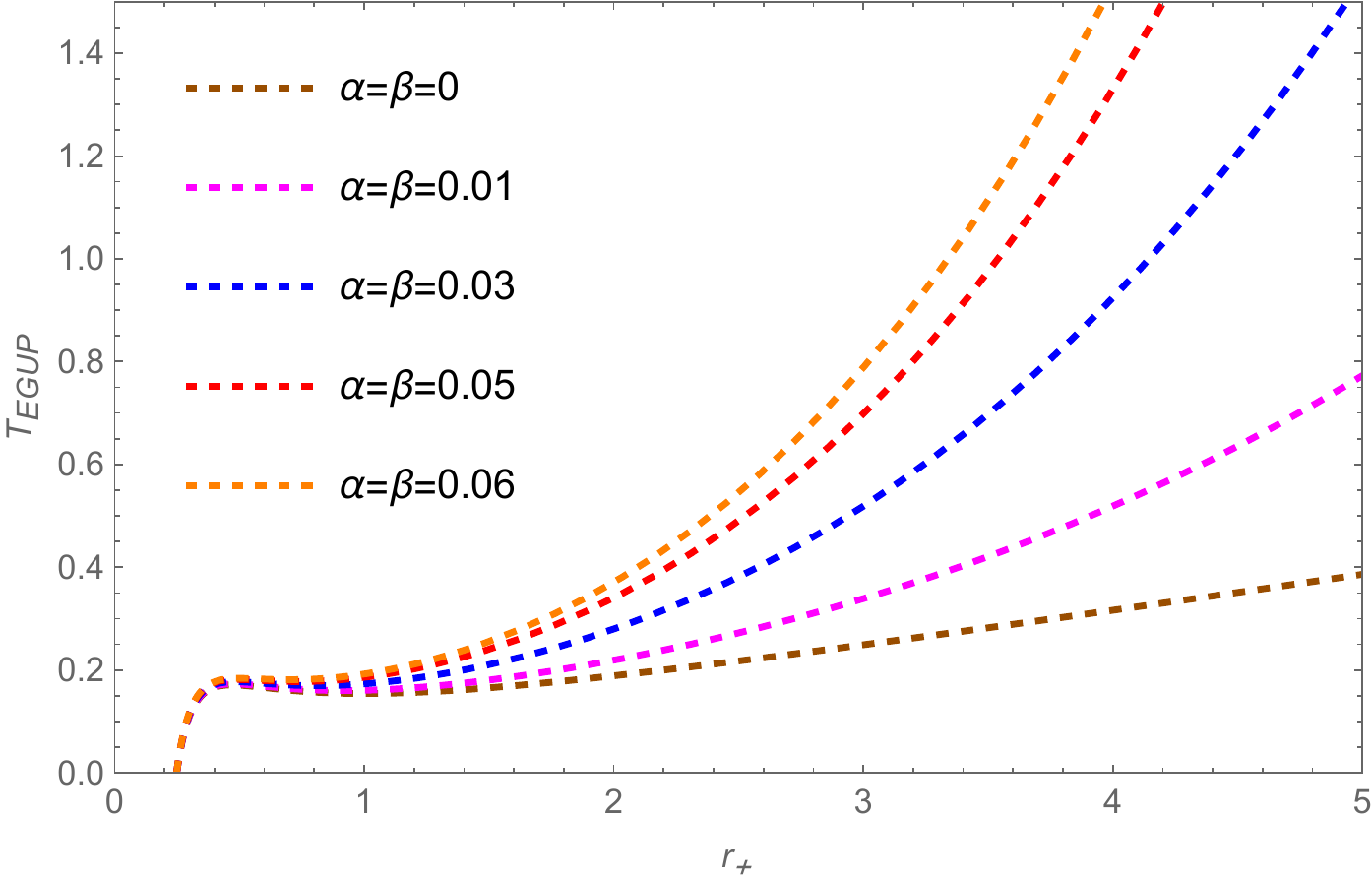}	
	\caption{The behavior of $T_{EGUP}$ vs. $r_+$ for different parameters $\alpha$ and $\beta$ $(l_p=L=1, l=\sqrt{\frac{3}{8 \pi \times  0.037}}, Q=0.3, \lambda =0.1)$.} 
	\label{liu1a}%
\end{figure}

Fig.~\ref{liu1a} shows that when the horizon radius remains constant, the black hole temperature increases with the increase in the EGUP parameters $\alpha$ and $\beta$. Moreover, the influence of the EGUP parameters is more obvious when the horizon radius is larger.

Next, we begin to derive the black hole heat capacity in the EGUP framework,
\begin{equation}\label{eq:E325}
C_{P}=\left(\frac{\partial M}{\partial T}\right)_{P}
\end{equation}

\begin{equation}\label{eq:E326}
C_{EGUP}=\frac{\pi\left(1-\frac{Q^{2}}{r_{+}^{2}}+\frac{\lambda}{r_{+}}+\frac{3 r_{+}^{2}}{l^{2}}\right)\left(1+\sqrt{1-\beta\left(\frac{\alpha}{L^{2}}+\frac{1}{4 r_{+}^{2}}\right) l_{p}^{2}}\right)^{2}}{2\left(2 \varepsilon r_{+}\left(\frac{2 Q^{2}}{r_{+}^{3}}-\frac{\lambda}{r_{+}^{2}}+\frac{6 r_{+}}{l^{2}}\right)+\left(1-\frac{Q^{2}}{r_{+}^{2}}+\frac{\lambda}{r_{+}}+\frac{3 r_{+}^{2}}{l^{2}}\right) \varepsilon_{0}\right)},
\end{equation}
with
\begin{equation}\label{eq:E327}
\varepsilon=\left(\frac{\alpha}{L^{2}}+\frac{1}{4 r_{+}^{2}}\right)\left(1+\sqrt{1-\beta\left(\frac{\alpha}{L^{2}}+\frac{1}{4 r_{+}^{2}}\right) l_{p}^{2}}\right),
\end{equation}
\begin{equation}\label{eq:E328}
\begin{split}
\varepsilon_{0}&=2 \varepsilon-\frac{\beta l_{p}^{2}\left(L^{2}+4 \alpha r_{+}^{2}\right)}{8 L^{2} r_{+}^{4} \sqrt{1+\beta\left(-\frac{\alpha}{L^{2}}-\frac{1}{4 r_{+}^{2}}\right) l_{p}^{2}}}\\&
-\frac{\left(1+\sqrt{1-\beta\left(\frac{\alpha}{L^{2}}+\frac{1}{4 r_{+}^{2}}\right) l_{p}^{2}}\right)}{r_{+}^{2}}.
\end{split}
\end{equation}
The black hole does not exchange radiation with the surrounding space when the heat capacity is zero, leading to the formation of a black hole remnant \cite{Chen:2022dap,Lin:2022doa,Lin:2022eix}. When the horizon radius satisfies $r_{ rem }(EGUP)= \frac{l_{p} L}{2} \sqrt{\frac{\beta}{L^{2}-\alpha \beta l_{p}^{2}}}$, substituting it into Eqs. (\ref{eq:E311}) and (\ref{eq:E35}) respectively, we obtain the remnant temperature and mass as follows:
\begin{equation}\label{eq:E329}
\begin{split}
T_{rem-EGUP}&=\frac{L}{\pi \beta l_{p}} \sqrt{\frac{\beta}{L^{2}-\alpha \beta l_{p}^{2}}}\Bigg(1+\frac{2 \lambda}{L l_{p} \sqrt{\frac{\beta}{L^{2}-\alpha \beta l_{p}^{2}}}}
\\&+\frac{3 L^{2} \beta l_{p}^{2}}{4 I^{2}\left(L^{2}-\alpha \beta l_{p}^{2}\right)}
-\frac{4 Q^{2}\left(L^{2}-\alpha \beta l_{p}^{2}\right)}{L^{2} \beta l_{p}^{2}}\Bigg)
\\&\times \left(1-\sqrt{1-\beta l_{p}^{2}\left(\frac{\alpha}{L^{2}}+\frac{L^{2}-\alpha \beta l_{p}^{2}}{L^{2} \beta l_{p}^{2}}\right)}\right),
\end{split}
\end{equation}

\begin{equation}\label{eq:E330}
\begin{split}
M_{rem-EGUP}&=\frac{1}{16}\Bigg(8 \lambda \ln_{}{\left ( \frac{L l_{p} }{2|\lambda|}\sqrt{\frac{\beta}{L^{2}-\alpha \beta l_{p}^{2}}} \right ) }  +\frac{16 Q^{2}}{L l_{p} \sqrt{\frac{\beta}{L^{2}-\alpha \beta 1_{p}^{2}}}}
\\&+4 L l_{p} \sqrt{\frac{\beta}{L^{2}-\alpha \beta l_{p}^{2}}}+\frac{L^{3} l_{p}^{3}}{l^{2}}\left(\frac{\beta}{L^{2}-\alpha \beta l_{p}^{2}}\right)^{3 / 2}\Bigg).
\end{split}
\end{equation}

When $\alpha=0$, the GUP-corrected heat capacity can be derived as 
\begin{equation}\label{eq:E331}
\begin{split}
C_{GUP}&=\frac{\pi}{2}\left(1-\frac{Q^2}{r_+^2}+\frac{\lambda }{r_+}+\frac{3 r_+^2}{l^2}\right)\left(1+\sqrt{1-\frac{\beta l_{p}^{2}}{4 r_{+}^{2}}}\right)^{2} \\& \times\Bigg(-\frac{\beta\left(-l^{2} Q^{2}+l^{2} \lambda r_{+}+l^{2} r_{+}^{2}+3 r_{+}^{4}\right) l_{p}^{2}}{4l^{2} r_{+}^{6} \sqrt{4-\frac{\beta l_{p}^{2}}{r_{+}^{2}}}}
\\&+\frac{\left(\frac{2 Q^{2}}{r_{+}^{3}}-\frac{\lambda}{r_{+}^{2}}+\frac{6 r_{+}}{l^{2}}\right)\left(1+\sqrt{1-\frac{\beta l_{p}^{2}}{4 r_{+}^{2}}}\right)}{2 r_{+}}
\\&-\frac{\left(1-\frac{Q^2}{r_+^2}+\frac{\lambda }{r_+}+\frac{3 r_+^2}{l^2}\right)\left(1+\sqrt{1-\frac{\beta l_{p}^{2}}{4 r_{+}^{2}}}\right)}{2 r_{+}^{2}}\Bigg)^{-1}.
\end{split}
\end{equation}
Similarly, if the horizon radius satisfies $r_{ rem }(GUP)=\frac{l_{p}}{2} \sqrt{\beta}$, we can derive the remnant temperature and mass in the GUP case, as shown in the following equations:
\begin{equation}\label{eq:E332}
\begin{split}
T_{rem-GUP}=\frac{1}{\pi \sqrt{\beta} l_{p}}\left(1-\frac{4 Q^{2}}{\beta l_{p}^{2}}+\frac{2 \lambda}{\sqrt{\beta} l_{p}}+\frac{3 \beta l_{p}^{2}}{4l^{2}}\right),
\end{split}
\end{equation}

\begin{equation}\label{eq:E333}
\begin{split}
M_{rem-GUP }=\frac{1}{16}\left(8 \lambda \ln_{}{\left ( \frac{\sqrt{\beta} l_{p}}{2|\lambda|} \right ) }  +\frac{16 Q^{2}+\beta l_{p}^{2}\left(4+\frac{\beta l_{p}^{2}}{l^{2}}\right)}{\sqrt{\beta} l_{p}}\right).
\end{split}
\end{equation}
For $\beta = 0$, the EUP-corrected heat capacity is given by 
\begin{equation}\label{eq:E334}
\begin{split}
C_{EUP}&=2 \pi\left(1-\frac{Q^{2}}{r_{+}^{2}}+\frac{\lambda}{r_{+}}+\frac{3 r_{+}^{2}}{l^{2}}\right)\Bigg(4 r_{+}\left(\frac{\alpha}{L^{2}}+\frac{1}{4 r_{+}^{2}}\right)
\\&\times \left(\frac{2 Q^{2}}{r_{+}^{3}}-\frac{\lambda}{r_{+}^{2}}+\frac{6 r_{+}}{l^{2}}\right)+\left(1-\frac{Q^{2}}{r_{+}^{2}}+\frac{\lambda}{r_{+}}+\frac{3 r_{+}^{2}}{l^{2}}\right)
\\&\times\left(4\left(\frac{\alpha}{L^{2}}+\frac{1}{4 r_{+}^{2}}\right)-\frac{2}{r_{+}^{2}}\right)\Bigg)^{-1}.
\end{split}
\end{equation}
When $\alpha=0$ and $\beta = 0$, the HUP-corrected heat capacity can be obtained as
\begin{equation}\label{eq:E335}
C_{HUP}=\frac{2 \pi\left(1-\frac{Q^{2}}{r_{+}^{2}}+\frac{\lambda}{r_{+}}+\frac{3 r_{+}^{2}}{l^{2}}\right)}{\frac{\frac{2 Q^{2}}{r_{+}^{3}}-\frac{\lambda}{r_{+}^{2}}+\frac{6 r_{+}}{l^{2}}}{r_{+}}-\frac{1-\frac{Q^{2}}{r_{+}^{2}}+\frac{\lambda}{r_{+}}+\frac{3 r_{+}^{2}}{l^{2}}}{r_{+}^{2}}}.
\end{equation}
For Eq. (\ref{eq:E335}), the heat capacity is reduced to $-2 \pi  r_+^2$ when $Q \rightarrow 0, \lambda \rightarrow 0, l \rightarrow \infty$, and the black hole becomes the Schwarzschild black hole.

\begin{figure}
	\centering 
	\includegraphics[width=0.45\textwidth]{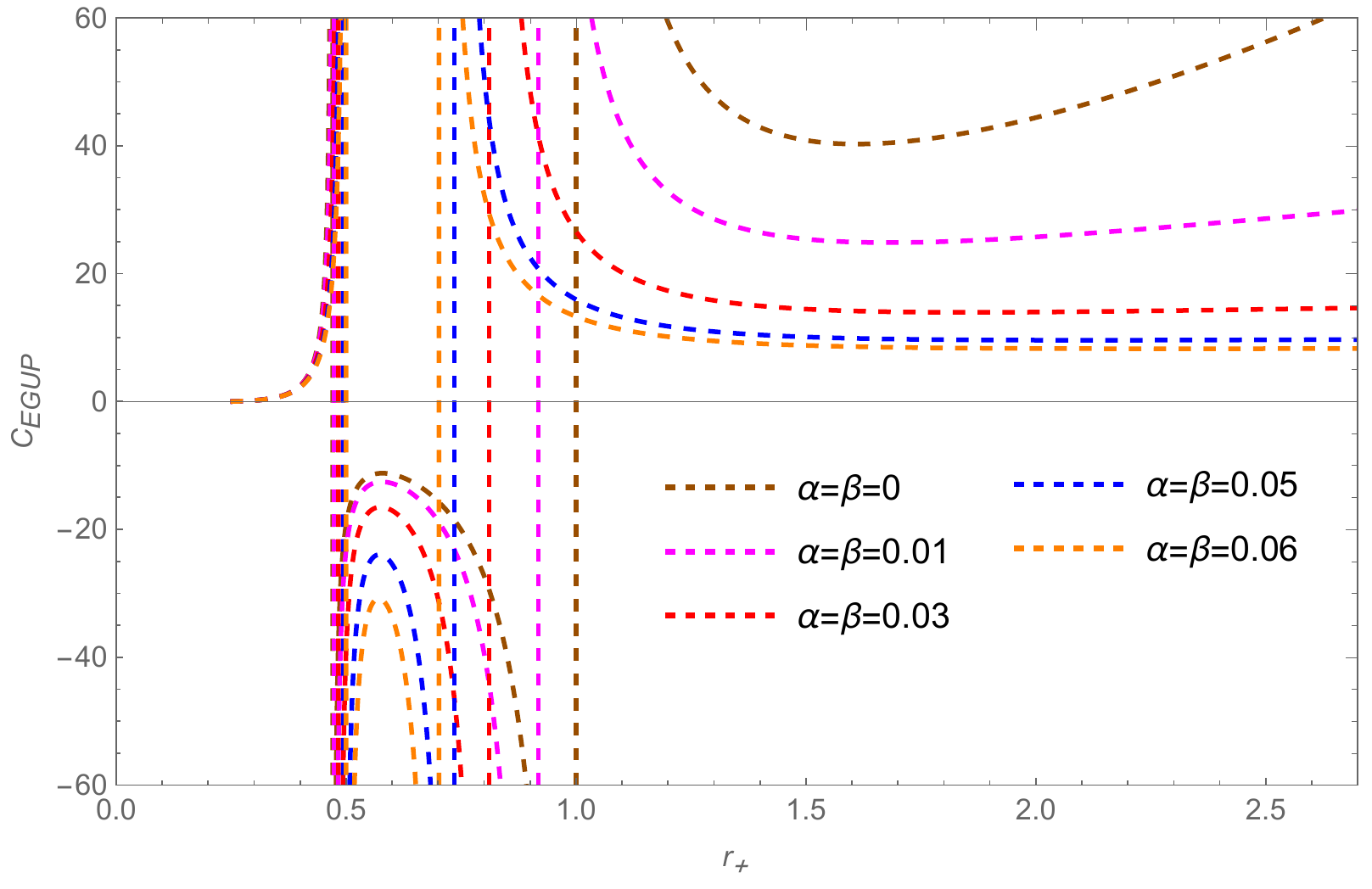}	
	\caption{The behavior of $C_{EGUP}$ vs. $r_+$ for different parameters $\alpha$ and $\beta$  $(l_p=L=1, l=\sqrt{\frac{3}{8 \pi \times  0.037}}, Q=0.3, \lambda =0.1)$.} 
	\label{liu2a}%
\end{figure}
Fig.~\ref{liu2a} shows that there are divergent points in heat capacity, indicating the occurrence of phase transition. We take the case of $\alpha=\beta=0.03$ as an example for discussion. When $r_+<0.483$ and $r_+>0.811$, the black hole heat capacity is positive, indicating that the black hole exists in a stable state, namely the small black hole (SBH) and the large black hole (LBH). In the range $0.483<r_+<0.811$, the black hole heat capacity is negative, corresponding to an unstable intermediate state, namely an intermediate black hole (IBH). Negative heat capacity also indicates the instability of the black hole thermodynamic properties. Considering the cases where $\alpha$ and $\beta$ take other values, we find that the negative heat capacity region decreases with increasing EGUP parameters, suggesting that the EGUP correction effect is beneficial to maintaining the thermodynamic stability of the black hole. For $r_+$ = 0.483 and $r_+$ = 0.81, the heat capacity diverges.  

Then, we need to deduce the black hole entropy after EGUP correction. First of all, from Eq. (\ref{eq:E37}), entropy is defined as
\begin{equation}\label{eq:E336}
S=\left(\int \frac{\partial M}{T}\right)_{Q, P, \lambda}=\int \frac{C \partial T}{T}.
\end{equation}
Substituting Eqs. (\ref{eq:E311}) and (\ref{eq:E326}) into Eq. (\ref{eq:E336}), we derive the EGUP-corrected entropy as follows:
\begin{equation}\label{eq:E337}
\begin{split}
 S_{EGUP}&=L^{2} \pi\Bigg(-L^{4}\left(L^{2}-4 \alpha\right) \beta l_{p}^{2}
 \\&\times\Bigg(2 \ln_{}{r_{+}}  
 +\ln_{}{\Bigg( L^{4}+4 L^{2} \alpha-\frac{\alpha \beta\left(L^{2}+4 \alpha r_{+}^{2}\right) l_{p}^{2}}{r_{+}^{2}}\Bigg) }  \Bigg) 
 \\&+4 r_{+}^{2}
 \bigg(2 L^{4}\left(L^{2}+4 \alpha\right)
 -L^{2} \alpha\left(L^{2}+12 \alpha\right) \beta l_{p}^{2}
 \\&+4 \alpha^{3} \beta^{2} l_{p}^{4}\bigg)\Bigg)
 \times \left(8\left(L^{4}+4 L^{2} \alpha-4 \alpha^{2} \beta I_{p}^{2}\right)^{2}\right)^{-1}.
\end{split}
\end{equation}
Similarly, we obtain the black hole entropy corrected by GUP and EUP respectively:
\begin{equation}\label{eq:E338}
\begin{split}
S_{G U P}=\pi r_{+}^{2}-\frac{1}{8} \pi \beta l_{p}^{2}\left(\ln_{}{L^{4}} +2\ln_{}{r_{+}} \right) ,
\end{split}
\end{equation}

\begin{equation}\label{eq:E339}
\begin{split}
S_{E U P}=\frac{L^{2} \pi r_{+}^{2}}{L^{2}+4 \alpha}.
\end{split}
\end{equation}
When $\alpha=\beta=0$, entropy can be reduced to the general expression $S=\pi r_{+}^{2}=A/4$.

\begin{figure}
	\centering 
	\includegraphics[width=0.45\textwidth]{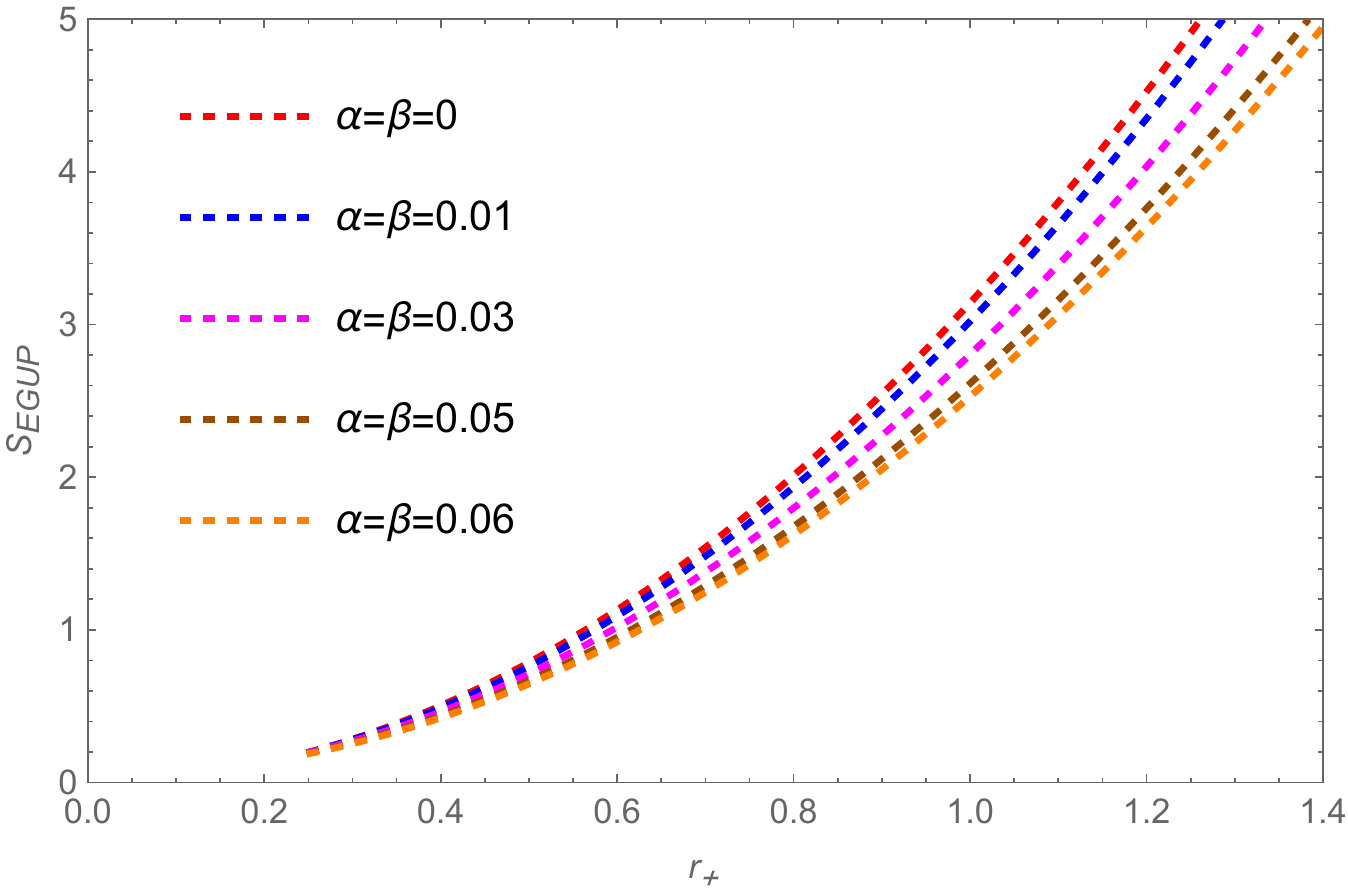}	
	\caption{The behavior of $S_{EGUP}$ vs. $r_+$ for different parameters $\alpha$ and $\beta$ $(l_p=L=1)$.} 
	\label{liu3a}%
\end{figure}
In Fig.~\ref{liu3a}, we observe that the corrected entropy is smaller than the uncorrected entropy. It can be found that the entropy decreases as the EGUP parameters $\alpha$ and $\beta$ increase while the horizon radius remains constant. The EGUP correction effect is more pronounced when the horizon radius is large.

\section{The P-v criticality in the EGUP framework}\label{IV}
From Eq. (\ref{eq:E311}), we can deduce the relationship between $P$ and $r_{+}$, which is expressed as 
\begin{equation}\label{eq:E340}
P=\frac{L^{2} Q^{2}-r_{+}\left(L^{2} \lambda+r_{+}\left(L^{2}-4 Q^{2} \alpha+r_{+}\left(4 \alpha\left(\lambda+r_{+}\right)-\omega\right)\right)\right)}{8 \pi r_{+}^{4}\left(L^{2}+4 \alpha r_{+}^{2}\right)},
\end{equation}
where
\begin{equation}\label{eq:E341}
\omega=2 L^{2} \pi T\left(1+\sqrt{1+\beta\left(-\frac{\alpha}{L^{2}}-\frac{1}{4 r_{+}^{2}}\right) l_{p}^{2}}\right).
\end{equation}
From Eq. (\ref{eq:E37}), the thermodynamic volume of a black hole can be given as 
\begin{equation}\label{eq:E342}
 V=\left(\frac{\partial M}{\partial P}\right)_{Q, S, \lambda}=\frac{4}{3} \pi r_{+}^{3}=\frac{\pi}{6} v^{3}.
\end{equation}
Here, the specific volume $v=2r_{+}l_p$, with $l_{p}=\sqrt{G \hbar / c^{3}}=1$ denoting the Planck length. The specific volume $v=2r_{+}$ is obtained by comparison with the equation of state for van der Waals fluids \cite{Kubiznak:2012wp}.\\
Therefore, based on Eqs. (\ref{eq:E340}) and (\ref{eq:E342}), we derive the equation of state for the RN-AdS black hole surrounded by PFDM as follows:
\begin{equation}\label{eq:E343}
\begin{split}
P&=\frac{1}{2 \pi}\Bigg(4 Q^{2} v^{2} \alpha-v^{3} \alpha(v+2 \lambda)+L^{2}\Bigg(4 Q^{2}+\pi T v^{3}-v^{2}-2 v \lambda
\\&+\pi T v^{3} \sqrt{1+\left(-\frac{1}{v^{2}}-\frac{\alpha}{L^{2}}\right) \beta l_{p}^{2}}\Bigg)\Bigg) \times\left(L^{2} v^{4}+v^{6} \alpha\right)^{-1}.
\end{split}
\end{equation}
The critical point condition for phase transition is given by
\begin{equation}\label{eq:E344}
\frac{\partial P}{\partial v}=\frac{\partial^{2} P}{\partial v^{2}}=0 .
\end{equation}
We consider $\lambda=0.1$, $L= l_p=1$ and $Q=0.3$. From Eq. (\ref{eq:E344}), the critical points where $\alpha$ and $\beta$ take different values can be obtained, as shown in the following table:

\begin{center}
\begin{table}[h]
\centering
\renewcommand{\arraystretch}{1.2}
\setlength{\tabcolsep}{0.28cm}
\begin{tabular}{|c|c|c|c|c|}
\hline

$\alpha=\beta$ & $P_c$ & $v_c$ & $T_c$ & $P_{c}v_{c}/T_c$ \\ 
\hline
0    & 0.06447 & 1.20000 & 0.19894 & 0.389   \\
0.01 & 0.06048 & 1.18729 & 0.19729 & 0.364  \\
0.03 & 0.05351 & 1.16803 & 0.19488 & 0.321    \\
0.05 & 0.04761 & 1.15386 & 0.19338 & 0.284 \\ 
\hline
\end{tabular}
\caption{The values of $P_c$, $v_c$, $T_c$ and $P_{c}v_{c}/T_c$ for different EGUP parameters $\alpha$ and $\beta$.}
\label{Td1}
\end{table}
\end{center}

\begin{figure}
  \centering
  \begin{subfigure}[b]{0.41\textwidth}
    \includegraphics[width=\textwidth]{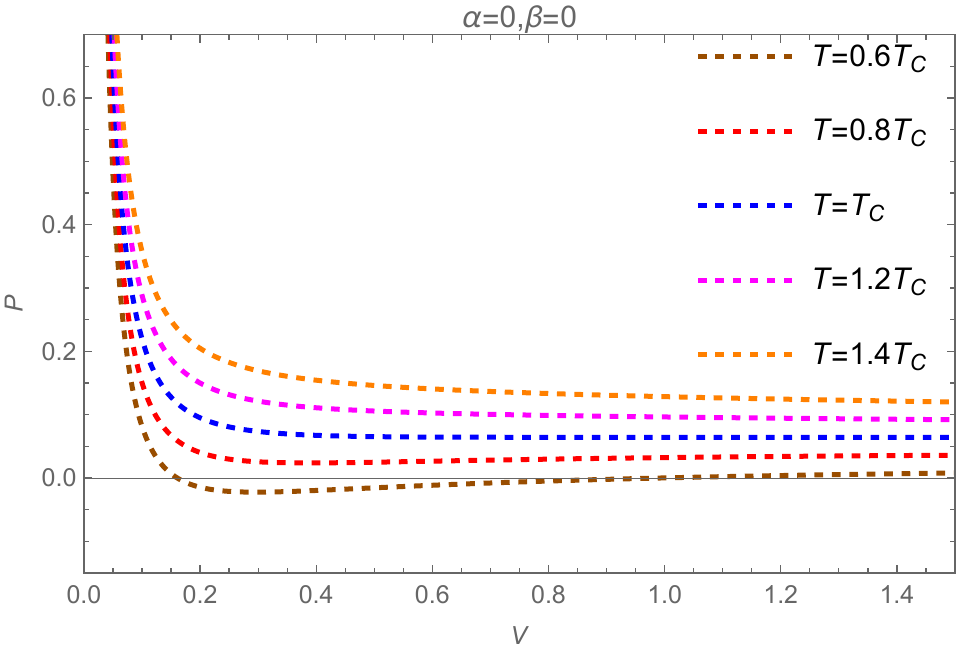}
    \caption{}
    \label{fig:4a}
  \end{subfigure}
  \quad
  \begin{subfigure}[b]{0.41\textwidth}
    \includegraphics[width=\textwidth]{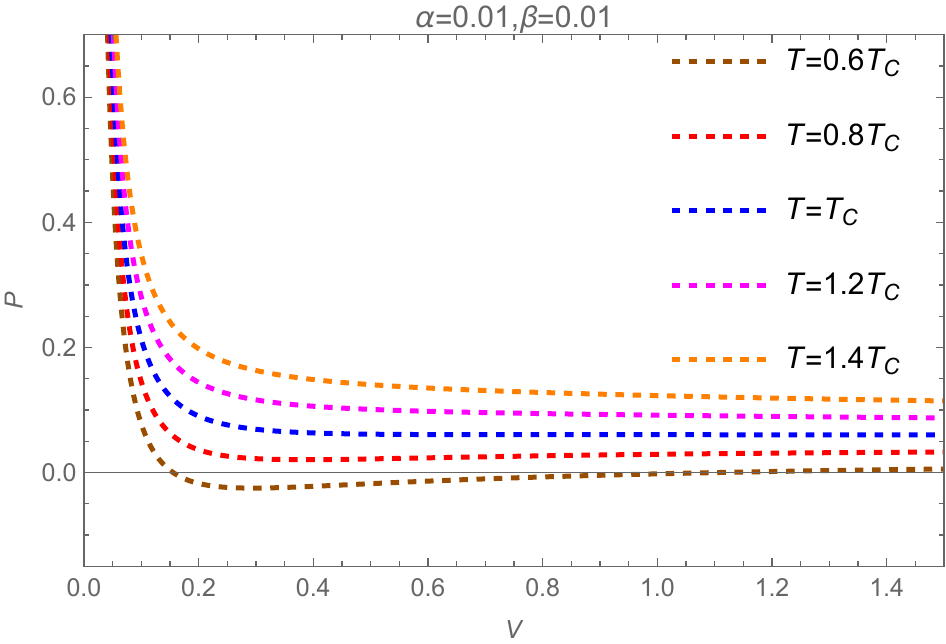}
    \caption{}
    \label{fig:4b}
  \end{subfigure}
  \quad
  \begin{subfigure}[b]{0.41\textwidth}
    \includegraphics[width=\textwidth]{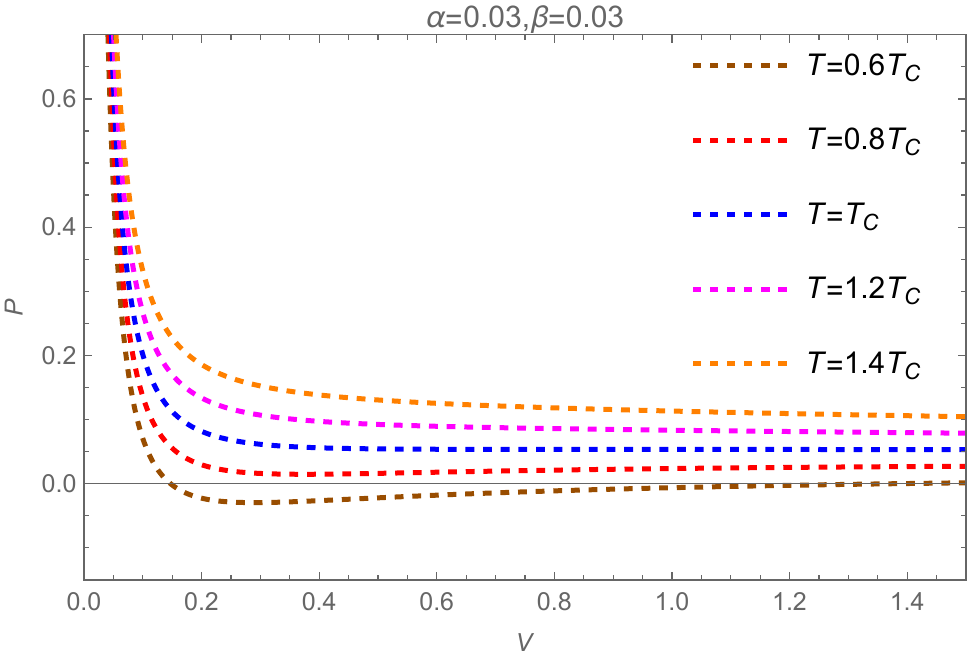}
    \caption{}
    \label{fig:4c}
  \end{subfigure}
  \quad
  \begin{subfigure}[b]{0.41\textwidth}
    \includegraphics[width=\textwidth]{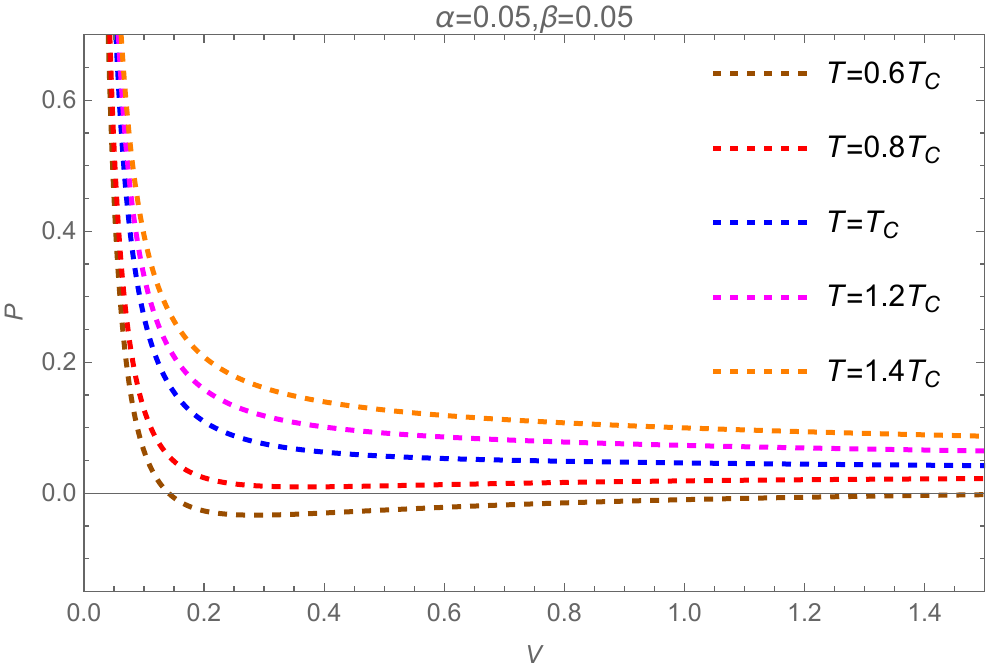}
    \caption{}
    \label{fig:4d}
  \end{subfigure}
  \caption{The behavior of $P$ vs. $v$ for different parameters $\alpha$ and $\beta$ ($l_p=L=1$, $Q=0.3$, $\lambda =0.1$, $T_c$ (see Table~\ref{Td1})).}
  \label{liu4a}
\end{figure}
It can be found that the P-v diagrams for different values of $\alpha$ and $\beta$ are similar in Fig.~\ref{liu4a}. The P-v isotherm exhibits behavior similar to that of an ideal gas when $T>T_c$. For $T<T_c$, the corresponding P-v isotherm exhibits a liquid-gas like phase transition behavior. When $T=T_c$, there is an inflection point on the isotherm, which can be obtained by Eq. (\ref{eq:E344}).

Finally, we derive the EGUP-corrected free energy using the following relationship,
\begin{equation}\label{eq:E345}
G = M - TS,
\end{equation}
\begin{equation}\label{eq:E346}
\begin{split}
G&=\frac{1}{24}\Bigg(12 \lambda \ln \frac{r_{+}}{|\lambda|}+\frac{12 Q^{2}}{r_{+}}+12 r_{+}+32 P \pi r_{+}^{3}
\\&+\frac{1}{\left(L^{4}+4 L^{2} \alpha-4 \alpha^{2} \beta l_{p}^{2}\right)^{2}} 3 L^{2} \pi T\Bigg(L ^ { 4 } ( L ^ { 2 } - 4 \alpha ) \beta \Bigg(2 \ln r_{+}
\\&+\ln \Bigg(L^{4}+4 L^{2} \alpha-\frac{\alpha \beta\left(L^{2}+4 \alpha r_{+}^{2}\right) l_{p}^{2}}{r_{+}^{2}}\Bigg)\Bigg) l_{p}^{2}
\\&-4 r_{+}^{2}\bigg(2 L^{4}\left(L^{2}+4 \alpha\right)
-L^{2} \alpha\left(L^{2}+12 \alpha\right) \beta l_{p}^{2}+4 \alpha^{3} \beta^{2} l_{p}^{4}\bigg)\Bigg)\Bigg).
\end{split}
\end{equation}

\begin{figure}
  \centering
  \begin{subfigure}[b]{0.41\textwidth}
    \includegraphics[width=\textwidth]{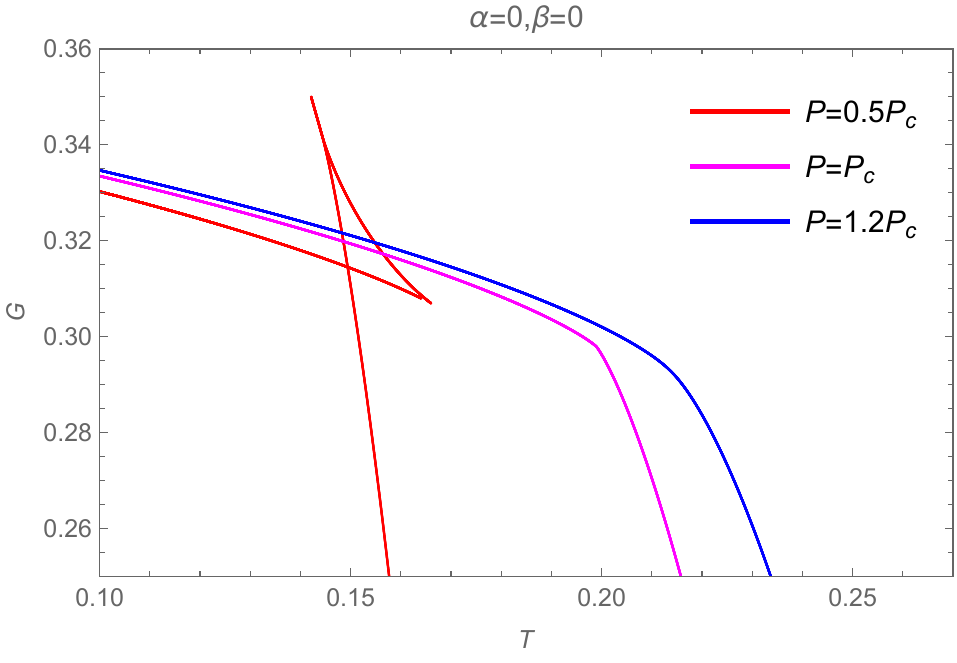}
    \caption{}
    \label{fig:5a}
  \end{subfigure}
  \quad
  \begin{subfigure}[b]{0.41\textwidth}
    \includegraphics[width=\textwidth]{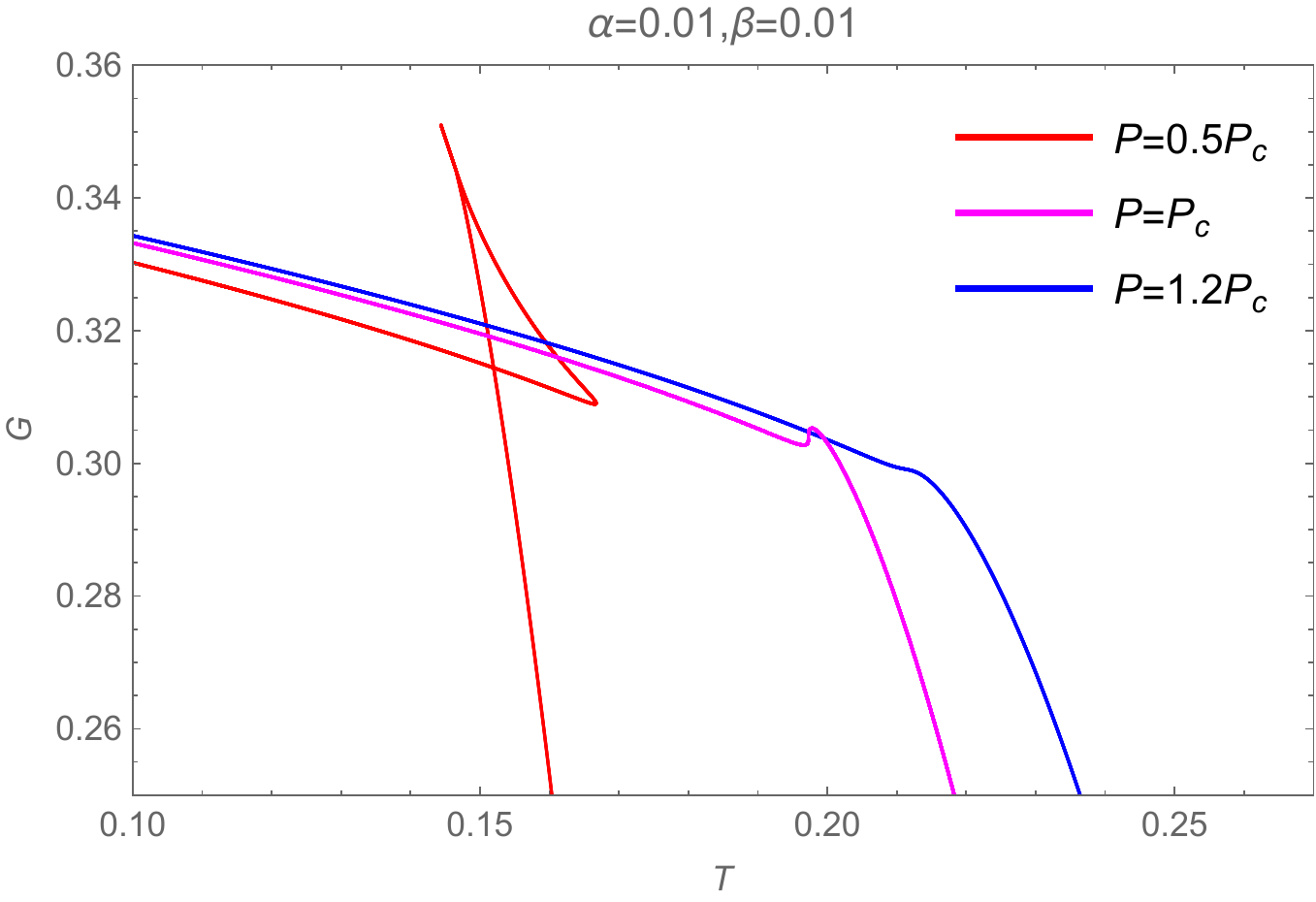}
    \caption{}
    \label{fig:5b}
  \end{subfigure}
  \quad
  \begin{subfigure}[b]{0.41\textwidth}
    \includegraphics[width=\textwidth]{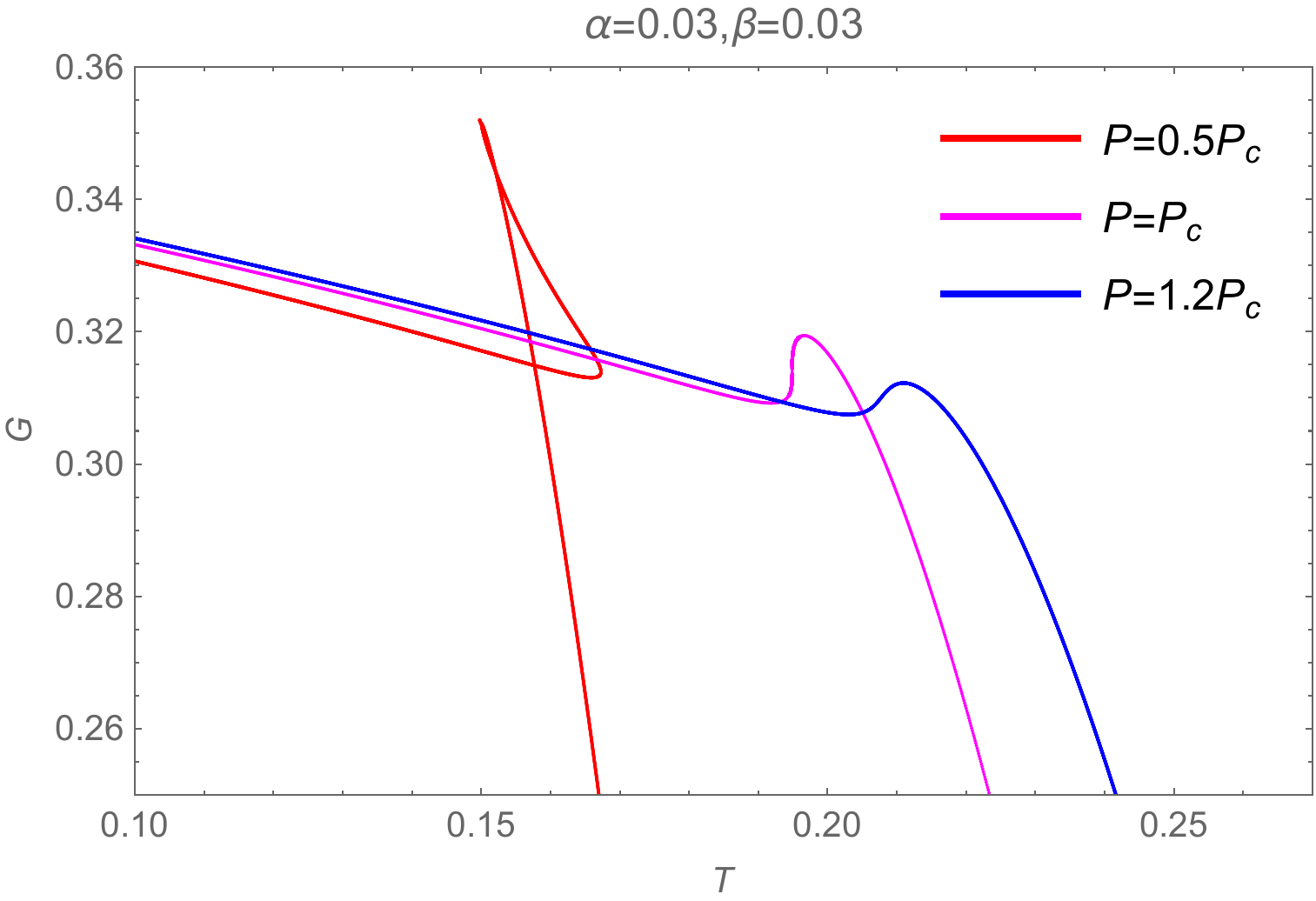}
    \caption{}
    \label{fig:5c}
  \end{subfigure}
  \quad
  \begin{subfigure}[b]{0.41\textwidth}
    \includegraphics[width=\textwidth]{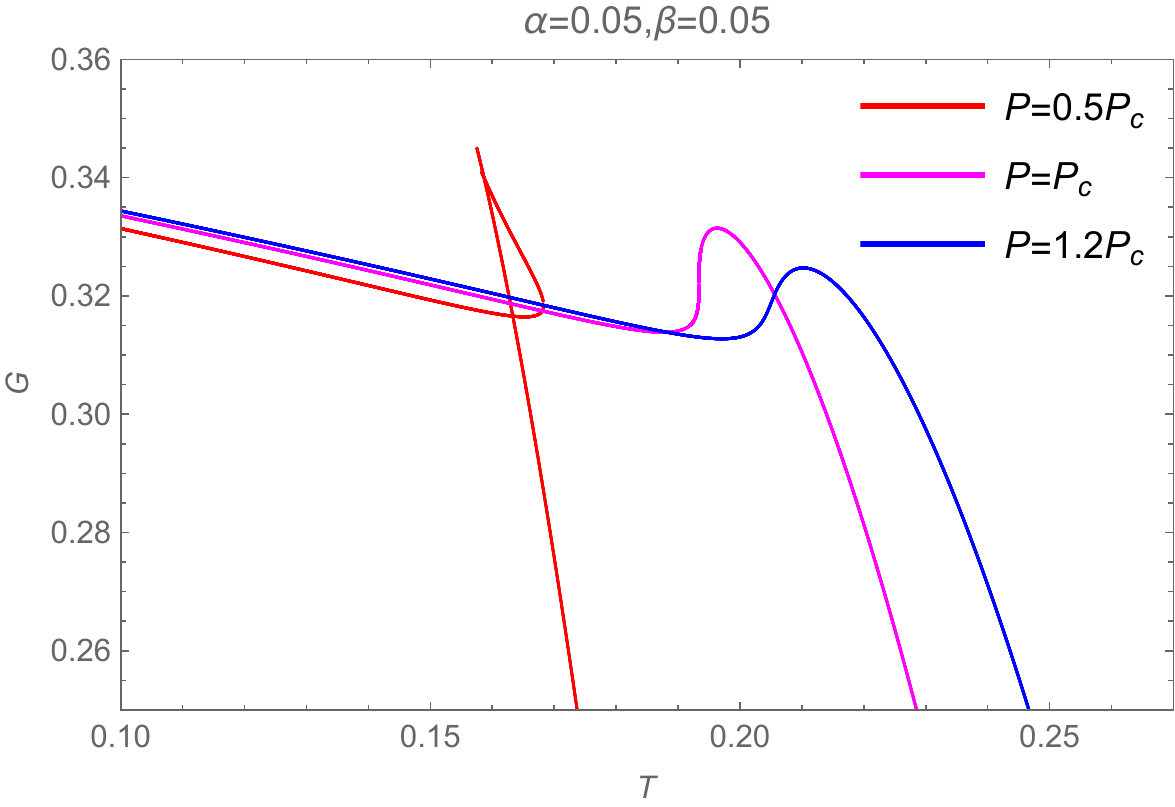}
    \caption{}
    \label{fig:5d}
  \end{subfigure}
  \caption{The behavior of $G$ vs. $T$ for different parameters $\alpha$ and $\beta$ ($l_p=L=1$, $Q=0.3$, $\lambda =0.1$, $P_c$ (see Table~\ref{Td1})).}
  \label{liu4b}
\end{figure}
Using the data from Table~\ref{Td1}, we plot in Fig.~\ref{liu4b} the relationship between the Gibbs free energy and the black hole temperature for different values of $\alpha$ and $\beta$. The four subfigures show that when $P<P_c$, ``swallowtail'' behavior occurs, indicating that the black hole undergoes a first-order phase transition. It can be found that the phase equilibrium point (i.e. the intersection of the ``swallowtail'' structure) changes slightly with increasing EGUP parameters. In Fig.~\ref{fig:5a}, a critical point appears when $P = P_c$. At this point, the first-order derivative of the free energy is continuous, while the second-order derivative is discontinuous, corresponding to the occurrence of a second order phase transition. For $P > P_c$, the curve becomes smooth, and both the first-order and second-order derivatives of the free energy are continuous, indicating that no phase transition occurs.

\begin{figure}[h]
  \centering
  \begin{subfigure}[b]{0.455\textwidth}
    \includegraphics[width=\textwidth]{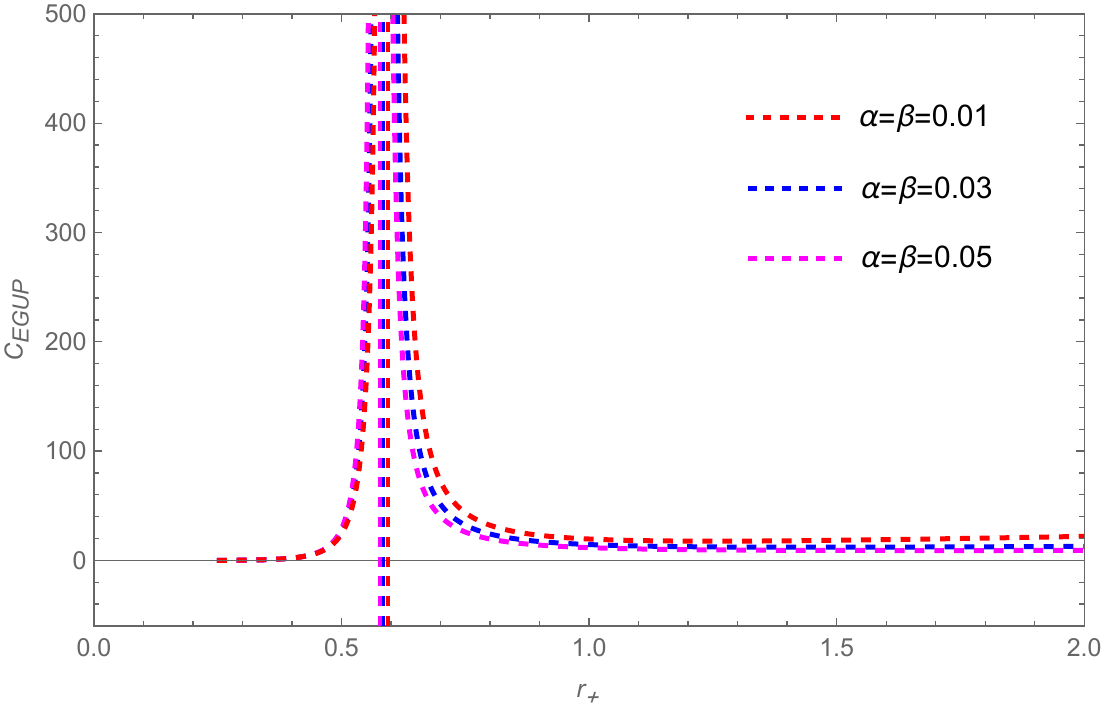}
    \caption{$P = P_c$}
    \label{fig:7a}
  \end{subfigure}
  \quad
  \begin{subfigure}[b]{0.455\textwidth}
    \includegraphics[width=\textwidth]{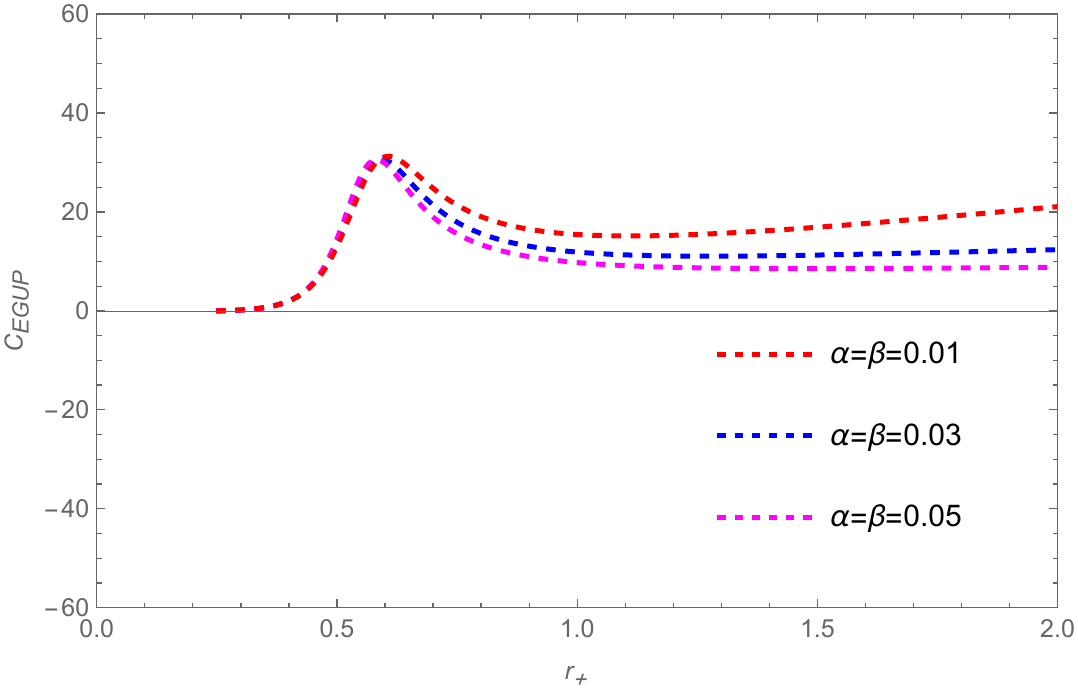}
    \caption{$P > P_c$}
    \label{fig:8b}
  \end{subfigure}
  \caption{Using $l=\sqrt{\frac{3}{8 \pi P}}$, we plot two distinct behaviors of $C_{EGUP}$ vs. $r_+$ for different parameters $\alpha$ and $\beta$ ($l_p=L=1$, $Q=0.3$, $\lambda=0.1$, $P_c$ (see Table~\ref{Td1})).}
  \label{liu4c}
\end{figure}
In order to analyze the cases of $P = P_c$ and $P > P_c$ in Fig.~\ref{fig:5b}, Fig.~\ref{fig:5c} and Fig.~\ref{fig:5d}, we plot the corresponding relationship between $C_{EGUP}$ and $r_+$ (Fig.~\ref{liu4c}). It can be observed that for $P = P_c$, the intermediate state disappears and a heat capacity divergence point appears, indicating a second-order phase transition from the SBH to the LBH. For $P > P_c$, the black hole undergoes no phase transition. Based on the above discussion, we can see that although the purple and blue curves in Fig.~\ref{liu4b} change as the EGUP parameters increase, the EGUP parameters actually have no substantial effect on the phase transition, and the phase transition characteristics are similar to the case of $\alpha$=$\beta$=0. The G-T plots for different EGUP parameter values show analogous properties, but as shown in Table~\ref{Td1}, the critical parameter varies significantly.

\section{Conclusions}\label{V}
In this work, we use the extended generalized uncertainty principle to study the thermodynamic properties and P-v criticality of the RN-AdS black hole surrounded by PFDM. To begin with, we derive the Hawking temperature after EGUP corrections. To ensure that the black hole temperature is a positive real value, we constrain the event horizon radius. Subsequently, we investigate the heat capacity and entropy after EGUP corrections and derive the remnant mass and temperature. The corrected heat capacity shows that, in the EGUP theoretical framework, the SBH and LBH are stable, while the IBH is unstable. The negative heat capacity region decreases as the EGUP parameters $\alpha$ and $\beta$ increase, suggesting that the modification is beneficial for maintaining the thermodynamic stability of black holes. Furthermore, the entropy decreases after the EGUP correction. Finally, we discuss the effect of EGUP on P-v criticality. Fig.~\ref{liu4b} shows that the black hole undergoes a first order phase transition when $P<P_c$, and the position of the phase equilibrium point changes slightly with increasing EGUP parameters. Then, combined with the analysis of Fig.~\ref{liu4c}, it can be found that for $P = P_c$ and $P > P_c$, the purple and blue curves in the G-T diagram are influenced by the EGUP parameters, but the phase transition characteristics of the two cases are similar to those of the uncorrected case.

\section*{Declaration of competing interest}
The authors declare that they have no known competing financial interests or personal relationships that could have appeared to influence the work reported in this paper.

\section*{Acknowledgements}
The work was supported by Yunnan Fundamental Research Projects (Grant No. 202301AS070029), and Yunnan Xingdian Talent Support Program-Young Talent Project.

\section*{Data availability}
No data was used for the research described in the article.




\end{document}